\documentstyle[11pt,aaspp4]{article}

\def\msolar{ {M_{\odot}} }

\slugcomment{To appear in The Astrophysical Journal Letters}
\lefthead{Briggs}
\righthead{Nearby LSB Galaxies}

\begin{document}

\title{Where are the Nearby Gas-Rich LSB Galaxies?}
\author{ F. H. Briggs}
\affil{Kapteyn Astronomical Institute, Postbus 800, 9700 AV Groningen, 
The Netherlands}

\lefthead{Briggs}
\righthead{Nearby LSB Galaxies}

\begin{abstract}
The Fisher-Tully ``10 Mpc Catalog of Late-Type Galaxies'' (FT, 1981) is 
remarkably complete.  Despite the considerable effort that has
been spent searching for  and cataloging low surface brightness 
(LSB) galaxies,
almost no new HI-rich galaxies have been added to
the volume to which the FT observations were both sensitive 
($d < 10 (M_{HI}/10^{8.2}\msolar)^{5/12}$ Mpc) and
intended to be complete ($d < 10$ Mpc).
It has not yet been demonstrated by the new surveys that
HI-rich LSB galaxies are not properly represented in this
catalog of nearby galaxies.  Although new optical
surveys are discovering heretofore uncatalogued
galaxies at greater distances than the
depth of the FT Catalog, they do not alter the completeness
of the FT Catalog for HI-rich LSB objects. Thus,
unless the galaxy population of the local volume is atypical,
there is at present no evidence for a significant, new population of 
gas-rich galaxies that has escaped recognition until now. 
\end{abstract}


\section{Introduction}
It has long been suspected that there could be a population of
galaxies  with surface brightness so low 
that the galaxies low contrast against
the dark sky would put them
below the detection levels of optical surveys (Arp 1965,
Disney 1976, Davies 1993).
More recently, new survey material and  computer-automated methods of 
analyzing survey plates have produced catalogs of galaxies specifically
selected for low surface brightness (Binggeli et al 1985, Schombert et al 1992, 
Davies et al 1988, Impey et al 1996).  Determination of the space density
of these objects can be very difficult, since the optical brightness for
the most extreme objects is so low that optical spectroscopy cannot provide
redshifts.  

Fortunately, the prototypical objects identified by several
LSB surveys are rich in neutral gas, allowing redshifts to be measured
in the 21 cm line (Longmore et al 1982,
Bothun et al 1985a, Schombert et al 1992).
The extreme of HI richness for LSB galaxies  includes objects whose
neutral hydrogen mass to optical luminosity ratio
$M_{HI}/L$  is well in excess of $1 M_{\odot}/L_{\odot}$.
At the other extreme is a class of dwarf LSB identified in the Virgo 
and Fornax clusters, that remains undetected in the 21cm line to very 
sensitive limits (${\sim}2{\times}10^6M_{\odot}$) (Bothun et al 1985b).  

This paper addresses the completeness of our knowledge of the gas-rich segment
of the nearby galaxy population.  In particular, it asks the question whether
recent surveys for LSB galaxies have added substantial
numbers of a ``new population''
to older catalogs of optically selected galaxies. The base
catalog is ``The 10 Mpc Catalog of Nearby Galaxies'' compiled by 
Fisher and Tully (1981; hereafter the FT Catalog). This catalog 
was designed to contain all late-type spiral galaxies and irregular
galaxies with declination above $-33^{\circ}$, Galactic latitude 
$|b_{II}| \geq 30^{\circ}$ within 10$h^{-1}$ Mpc of the Milky Way
($h=H_o/100$ km~s$^{-1}$). 
The catalog was constructed from a complete inspection of the PSS
using morphological classification to select galaxies expected
to have redshifts less than 2000~km~s$^{-1}$.  
The catalog was supplemented by including all galaxies except
ellipticals from the UGC (Nilson 1973) with diameter 3$'$ or greater
and all late-type systems (Sd to Im) with diameter $>2'$.
In total, 1787 galaxies were  observed 
in the 21cm line to a uniform flux density level, with detection reported
for about 2/3 of the sample. The majority of the 
detections lie beyond 10 Mpc.

There are several newer catalogs of optically selected galaxies
for which 21cm line detection has provided an efficient method for
redshift determination.   Here we choose
to parameterize the population by  HI mass, since the HI measurement provides
both a 
distance indicator and a measure of each galaxy's detectability.  The
detectability function translates into a volume sensitivity that Briggs
and Rao (1993) and Briggs (1997) have used to estimate the spatial
density of gas-rich galaxies.
Since members of the newer catalogs were observed at higher sensitivity
than the FT observations (using the Arecibo Telescope, for example), 
examples of familiar nearby galaxies can be now detected at much
larger distances, and thus they can be drawn from a larger volume, as
will be shown to be the case for the new catalogs.

\section{Analysis}

The comparison of the new samples with 
the FT Catalog takes place in Figure 1, where each galaxy
is registered as a point with its distance along the vertical axis
and its measured HI mass along the horizontal.  This type of plot 
enables a clear view of the detection limits for galaxies of different
HI mass and indicates from which volumes the different samples are drawn.

The HI mass $M_{HI}$
is a derived quantity that depends on each galaxy's distance; 
the effects of  errors in distance and flux measurement on the location
of the points are illustrated by the exaggerated
error bars  in the upper right panel. A distance
error will move a point diagonally in 
these diagrams. 
Measurement  errors for redshift
are typically  negligible once a galaxy is reliably detected,
and it is  peculiar velocity that dominates in displacing
the galaxies along the diagonal error bar; since
the displacement is very nearly parallel to the detection boundary, even
very large uncertainties in distance fail to move galaxies across
the the detection boundary.
Flux errors  cause  horizontal displacements of typically less than
20 percent.  
 
The FT Catalog, plotted in the upper left, has a clear detection HI
limit that appears as a diagonal boundary between the area at the
lower right, which is heavily populated, and the upper  left, 
where only a few points
spill over the boundary.  A heavy line with slope $d_c \propto M_{HI}^{5/12}$
marks the expected form of the boundary, with the recognition that more
massive galaxies (with larger HI mass) also rotate faster, spreading their
signals into broader profiles of lower amplitude that are harder to detect. 
If all galaxies had the same velocity spread, $d_c$ would be
$\propto M_{HI}^{1/2}$. The effect of profile width,
$\Delta V$, on detectability
was explored in greater detail by Briggs \& Rao (1993) and is discussed
only briefly here. The noise level, $\sigma$,
in a spectrum that has been optimally
smoothed to match the profile is $\sigma \propto 1/\sqrt{\Delta V}$.
Since $M_{HI}= 2\times 10^5 d_{Mpc}^2 \int S_{Jy} dV_{km/s}\msolar$,
the minimum detectable HI mass is 
$M_{HI} \propto 5\sigma \Delta V d^2 \propto d^2\sqrt{\Delta V}$,
if the minimum detectable profile is modeled as a rectangle of height
in flux density $\Delta S = 5\sigma$ and width $\Delta V$.
A sort of HI Tully-Fisher relation has 
$\Delta V \propto M_{HI}^{1/3}\sin i$ for galaxies with inclination $i$
relative to the plane of the sky,
leading to the result that $d_c \propto M_{HI}^{5/12}\sin^{-1/4}i$.
In fact, enough
information exists in the FT catalog to make a first order correction
for inclination, which does indeed sharpen the detection boundary (Briggs
\& Rao 1993). On the other hand, since the 
inclination and size information is less well determined or not provided
for all the samples and the $\sin^{-1/4}i$ factor is substantially
different from unity for a only small fraction of a randomly oriented sample,
we choose here to not take advantage of the information on
inclination for the few catalogs in which it is given. 

Other features in the distribution of
FT points in Figure 1 are: (1) At the largest distances and
largest masses, there is a fairly abrupt fall off in the density of points;
this arises from the limited spectrometer bandwidth available at the time
of the FT survey, which concentrated on the redshift range $-400$ to
$+3000$ km~s$^{-1}$, but the lack of coverage at higher redshift
was not of concern since the sample was intended
for completeness at $d < 10$~Mpc. (2) There is a concentration of points
in a line at $\log 10.8$ Mpc $=1.03$ due to the assumption that galaxies
within 6$^{\circ}$ of the core of the Virgo cluster lie at this distance.
(3) There are four points that lie on a horizontal
line at 1 Mpc, due to assigning
a distance of 1 Mpc to any galaxy with velocity less than 100 km~s$^{-1}$
(measured with respect to the centroid of the Local Group).

Distances and masses for the newer samples plotted in Figure 1
are computed in the same manner as those for the FT galaxies,
assuming a Hubble Constant of 100~km~s$^{-1}$Mpc$^{-1}$.

The detection boundary prevents the FT 10 Mpc catalog from attaining
completeness to the full 10 Mpc depth
for $M_{HI} < 10^{8.2}$M$_{\odot}$.
However, the detection boundary
is well defined and well understood, and the catalog can be tested
for ``completeness'' within this boundary. Furthermore, the
volume within which a given HI mass is detectable can be calculated,
allowing the catalog to be used to
estimate the space density of low $M_{HI}$ objects that meet
the late type morphological selection (Briggs 1997).
The one-third of the FT sample that was undetected in the original catalog
lies above the line by definition.

The detection boundary for the FT Catalog is replotted in the other 
panels of Figure 1 for comparison with the other samples.  For most of the
other samples, the high sensitivity of the Arecibo Telescope
detects galaxies far beyond the FT sensitivity boundary.
On the other hand, the most striking
result is that the new surveys add almost no galaxies to the volume of
space where the FT Catalog was sensitive.  
Table 1 lists $N_{CZ}$ and $N_{SZ}$, the numbers of new galaxies in each
catalog falling in the FT Zone of Completeness (CZ) and the Zone of
Sensitivity (SZ), respectively. The CZ and SZ are defined in Figure 1.
(Note that the $N_{CZ}$ galaxies in the CZ are
a subset of those counted in $N_{SZ}$.) Since there is overlap
between catalogs, construction of these counts
requires exclusion of galaxies in the
Schneider et al (1990) ``Catalog of UGC Dwarfs and LSB Galaxies'' that were
already listed in the FT Catalog. Furthermore, there are galaxies in the 
Southern Extreme Late-Type Galaxies Catalog (Matthews \& Gallagher 1996,
Matthews et al 1995, Gallagher et al 1995) that fall at declinations or
Galactic latitudes outside the region covered by the FT Catalog,
and these are also excluded from the $N_{CZ}$ and $N_{SC}$ listed
under Mathews et al in the table. The
Schombert et al (1992) ``Catalog of LSB Galaxies'' adds no new members to the
CZ and one to the SZ, 
while the automated selection of LSB objects in the
catalog of Impey et al (1996) yields one detection in the CZ and one
galaxy in the SZ at redshift less than 3000~km~s$^{-1}$.
In fact, 
the new surveys cover smaller solid angles $\Omega$ 
of sky than the FT Catalog, 
so that a weighting factor, $\Omega_{FT}/\Omega$, 
has been applied to the counts for the new catalogs (column (4) in the
Table) for a fair comparison with the FT numbers.
The very few new galaxies that are added to either the CZ or the SZ by
any sample are located
close to the detection boundary, as though they represent a simple
incompleteness in the FT Catalog due to HI detection sensitivity. 

The recent literature has other
new catalogs of objects with a variety of selection criteria
in addition to the catalogs plotted in Figure 1. For example,
the six prototypes of the ``dwarf spiral''  class presented by Schombert
et al (1995) lie at redshifts from 3200 to 6000 km~s$^{-1}$ -- entirely outside
the FT volume.  The few dwarf galaxies from the sample
compiled by Eder et al (1989, 1996)
that fall within the SZ are previously cataloged objects with UGC numbers,
and thus they also make no new addition to the inventory of nearby galaxies.

The few new galaxies added to the CZ (within $10h^{-1}$ Mpc) 
are low in HI mass, and thus they add only a few percent
to the HI mass content of the nearby Universe.  Estimates of  
average HI densities are
listed in Table 1, resulting from   summations over the members
of each sample falling in the CZ or SZ:
$$\rho_{HI} = \sum\frac{M_{HI}}{V_m}$$
where $V_m = \Omega d_L^3/3$ for $d_L=d_c$ in the sensitivity bounded
mass ranges and $d_L=10$ Mpc or 30 Mpc in the large mass ranges of
the CZ or SZ samples respectively.  
The density of ${\sim}10^8\msolar$ is higher than most estimates 
(Fall \& Pei 1989, Rao \& Briggs 1993, Zwaan et al 1997) due to the
bias caused by drawing the FT Catalog from the Local Supercluster
(Felton 1977).
It is clear from Table 1 that the
new samples add a small number of galaxies and a tiny
amount of HI mass to the volume where FT Catalog is sensitive.

\section{Discussion}

If the recent surveys for low surface brightness galaxies were
finding populations that are both
(1) new and (2) cosmologically dominant
repositories of matter, then the new detections
should outnumber the old ones in any large volume.  Instead, the
new catalogs fill more distant 
shells in the $d$--$M_{HI}$ space of Figure 1.
Progressing from the FT diameter limit $\Theta_L{\approx}2'$, 
through the $1'$ limit
of the UGC to the $30''$ limit (at fainter isophotes) 
for Schombert et al (1992), 
Figure 1 shows that the catalog members lie at successively larger distance
as the angular diameter limit is reduced.
Within the Impey et al (1996) catalog ($\Theta_L{\approx}30''$ 
for the bulk of the catalog), there is a gradient across the shell of
HI detections, with the previously catalogued members of their sample
falling at smaller distance and the new objects at larger distance.
This behavior signifies that the new surveys are tending to select more
distant examples of familiar types of galaxies, without adding
to the density of galaxies or to the mass content of the Universe.
The implication of Figure 1
is that the types of galaxies selected for their LSB
properties by Schombert et al (1992) and by Impey et al (1996) must already be
fairly represented by nearby, previously cataloged examples.

Why might there already be LSB galaxies in the historical catalogs?
Much of the answer may be that the new catalogs look for 
LSB ``disks'' whose extrapolated central surface brightness  
lies below a set threshold and whose angular extent exceeds a set 
diameter limit.  The extrapolation may lie beneath a luminous
region at the
center of the galaxy or patchy emission
that is  bright enough to attract the attention
of the compilers of the earlier surveys.  It may be the case that a
fraction of the new LSB galaxies would be given other classifications if
they were located close by.
These kinds of nearby galaxies
may have never been studied with large format detectors and observing
techniques that would be sensitive to very extended LSB light without
removing it during a ``sky subtraction'' step in the data reduction.

Is it possible that a gas-rich 
population of still lower optical surface brightness
exists that could escape inclusion in these galaxy samples?  
Since the prototypical
LSB galaxies that are most broadly represented in the
new samples are rich in neutral gas (Schombert et al 1992), sky
surveys in the 21cm line make effective tests of the existence of a gas-rich,
extreme LSB population. A number of 
``blind surveys'' in the 21cm line have produced HI-selected
galaxy samples with  significant numbers of members
(Lo \& Sargent 1979,
Kerr \& Henning 1987, Henning 1995,  
Weinberg et al 1991, Szomoru et al 1994, Sorar 1994, 
Spitzak 1996, Zwaan et al 1997, Schneider 1997).
Subsequent optical follow up  has found optical counterparts to more than
99 percent of the HI selected objects; the only exceptions appear at the 
very faint end of the mass distribution, where they make only small 
contributions to the integral mass density.  The HI selected objects
are apparently drawn from the same population of extragalactic system that is
already cataloged by optical selection.  Reported ``intergalactic HI clouds''
have so far been subsequently associated with visible galaxies, either
as tidal remnants (e.g. Chengalur et al 1995) or as a bound ring
(Schneider 1989).

\section{Conclusion}

There is no evidence so far that new catalogs of LSB galaxies are 
identifying a new population.  Clearly, traditional catalogs such
as the UGC, CGCG (Zwicky et al 1961), etc have completeness limits,
and new surveys
that are sensitive to lower surface brightness or accept galaxies
of smaller angular diameters will always add distant 
galaxies that have not already been catalogued.  
Indeed, the new catalogs appear to be adding detections in successively
more distant shells in the $d$ {\it vs.} $M_{HI}$ diagram.
The new surveys are showing that
the catalog of Fisher and Tully (1981) is
remarkably complete for the nearby volume, and the implication is
that it must already
contain a balanced representation of gas-rich LSB galaxies.

\acknowledgements

P. Sackett, R. Sancisi, M. Zwaan, and C. Impey 
have contributed useful comments and criticism to this work.
The author is also grateful to James Schombert and Otto Richter for 
providing computer
files containing the Low Surface Brightness Galaxy Catalog and the
Huchtmeier-Richter Catalog of HI Observations of Galaxies, respectively.
This work was supported in part by NSF Grant AST 91-19930.

\newpage

\begin{figure}
\plotone{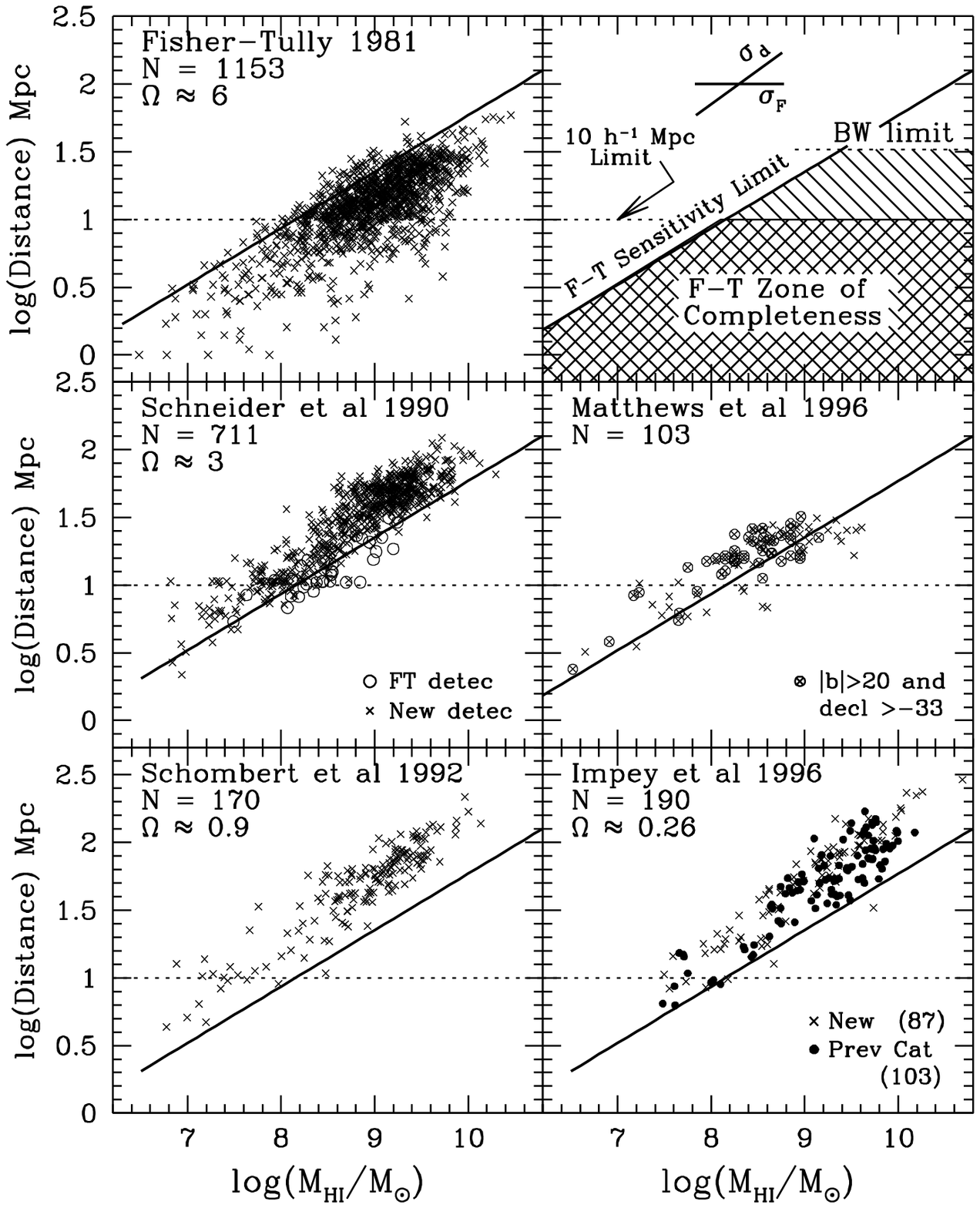}
\caption{
Depths of the HI-rich galaxy samples. 
For five galaxy samples, the galaxies detected in HI are 
plotted at their measured distance as a function of their HI mass. 
An estimate for the HI detection limit for the Fisher-Tully (1981)
sample $d_c \propto M_{HI}^{5/12}$
is drawn as a heavy solid line in all six panels.
A horizontal dotted line in each panel 
indicates 10 Mpc ($H_o = 100$ km~s$^{-1}$~Mpc$^{-1}$).
The number of galaxies detected, $N$, and the
survey solid angle, $\Omega$,  for each catalog are indicated. A subset
of the Schneider et al catalog was already detected by
FT, and these are indicated by open circles.
Points representing galaxies in the Matthews-Gallagher samples 
falling at $|b|>20^{\circ}$ and declination $>-33^{\circ}$ are enclosed 
in circles. 
A total of 103 of the 190
galaxies detected in HI in the Impey et al (1996) survey had been
previously catalogued.
The {\it top 
right panel} shows a crosshatched region where the FT Catalog is nominally
complete (the Completeness Zone, CZ); 
the FT Sensitivity Zone (SZ) is shaded and consists of the entire region
below the F-T sensitivity limit (including the CZ)
and within the $d < 30 h^{-1}$Mpc bandwidth (BW) 
limited depth of the FT survey. 
The effects of a factor of 3 error in measuring flux
and a factor of $\sqrt 3$ in determining the distance are indicated.
}
\end{figure}

\begin{deluxetable}{lrrrrrr}
\tablewidth{425.0pt}
\tablenum{1}
\tablecaption{Detections in the F-T Zones of Completeness and Sensitivity}
\tablehead{
\colhead{Catalog} & \colhead{$\Omega_s$} &  
\colhead{$N_{CZ}$} &\colhead{$N_{SZ}$} & $\frac{\Omega_{FT}}{\Omega_s}N_{CZ}$ &
\colhead{$\rho_{HI}[CZ]$} &\colhead{$\rho_{HI}[SZ]$} 
\nl
 & \colhead{Sterad} &
\colhead{ } &\colhead{ } &  &
\colhead{M$_{\odot}$Mpc$^{-3}$} &\colhead{M$_{\odot}$Mpc$^{-3}$}
\nl
\colhead{(1)} & \colhead{(2)} & \colhead{(3)} & \colhead{(4)} & 
\colhead{(5)} & \colhead{(6)} & \colhead{(7)}  
}

\startdata
Fisher-Tully 1981  & ${\sim}6$~ & 315  &  1017 & 315 ~ &10.8$\times$10$^7$ & 5.8$\times$10$^7$ \nl
%~ ~ 1981 \nl
              \nl
Schneider et al 1990& ${\sim}3$~ & 12  & 28  &  24 ~ & 2.4$\times$10$^6$ & 2.7$\times$10$^6$ \nl
%~ ~ 1990 &  \nl
            \nl
Schombert et al 1992& ${\sim}0.9~ $ & 0   & 1 &0 ~ &0 ~ ~~ &3.8$\times$10$^5$   \nl
%~ ~ 1992 &  \nl 
           \nl
Matthews et al 1996& $<1~ $ & 1 & 8 & $>6$ ~ & $>5\times$10$^5$&$>2.1\times$10$^6$ \nl
%~ ~ 1996   \nl 
           \nl
Impey et al 1996& 0.26~ & 1 & 2 & 23 ~ & 1.7$\times$10$^6$ & 4.0$\times$10$^6$   \nl
%~ ~ 1996   \nl 
\enddata

\tablecomments{Columns: (1) catalog by authors, 
(2) survey solid angle (steradians),
(3) $N_{CZ}$, number of new galaxies within FT zone of completeness,
(4) $N_{SZ}$, number of new galaxies within FT zone of sensitivity,
(5) $N_{CZ}$ from column 2, scaled to the solid angle, $\Omega_{FT}$, of
the FT Catalog,
(6) average HI mass density within the FT zone of completeness ($d < 10 h^{-1}$
Mpc) contributed by each survey,
(7) average HI mass density within the FT zone of sensitivity from each
survey.
}

\end{deluxetable}

\end{document}